\documentclass{aa}
\usepackage[dvips]{graphicx}
\usepackage{psfig}

\newcommand{\lapp}{\mbox{\raisebox{-0.3em}{$\stackrel{\textstyle <}{\sim}$}}}

\begin{document}

\title{Jet propagation and the asymmetries of CSS radio sources}

\author{S.\ Jeyakumar \inst{1,2} \and Paul J.\ Wiita \inst{3} \and
D.\ J.\ Saikia \inst{4} \and  Jagbir S.\ Hooda \inst{5} }

\titlerunning{Propagation of jets and the asymmetries of CSS sources}

\offprints{S. Jeyakumar, sjk@rri.res.in}

\institute{
Physikalisches Institut, Universit\"{a}t zu K\"{o}ln, Z\"{u}lpicher Str. 77, 
50937 K\"{o}ln, Germany
\and
Raman Research Institute, C.\ V.\ Raman Road, Bangalore 560080, India
\and
Department of Physics \& Astronomy,
Georgia State University,  P.O.\ Box 4106, Atlanta GA 30302-4106, USA
\and
National Centre for Radio Astrophysics,
TIFR, Post Bag No. 3, Ganeshkhind, Pune 411 007, India
\and
Software Engineering, E*trade Financial Corporation, 4500 Bohannon Drive, Menlo Park, CA 94025, USA
}

\date{Received 0000; Accepted 0000}

\abstract{
As Compact Steep Spectrum radio sources have been shown to be more
asymmetrical than larger sources of similar powers, there is a high
probability that they interact with an asymmetric medium in the
central regions of the host elliptical galaxy.  We consider a simple
analytical model of the propagation of radio jets through a
reasonable asymmetric environment and show that they can yield the
range of arm-length and luminosity asymmetries that have been observed.
We then generalize this to allow for the effects of orientation,
and quantify the substantial enhancements of the asymmetries that
can be produced in this fashion.  We present two-dimensional
and three-dimensional simulations of jets propagating through
multi-phase media and note that the results from the simulations are
also broadly consistent with the observations.

}

\maketitle

\section{Introduction}
Compact Steep-spectrum Sources (CSSs), defined to be $\lapp$20 kpc in size
(for $H_0$=100 km s$^{-1}$ Mpc$^{-1}$ and $q_0$ = 0.0) and having a steep
high-frequency radio spectrum  ($\alpha \geq 0.5$ where  S$_{\nu} \propto \nu^{-\alpha}$),
have received a great deal of attention in recent years.
The general consensus is that these are largely young sources seen at an
early stage of their evolution (e.g., Carvalho 1985; Fanti et al.\ 1995; 
Readhead et al. 1996a,b; see O'Dea 1998 for a review).  
Related to the CSSs are the very small ($\leq 1$ kpc) 
Gigahertz-Peaked-Spectrum (GPS) sources (e.g., Gopal-Krishna, Patnaik \& Steppe 1983; 
Stanghellini et al.\ 1998).  
The smallest double-lobed radio sources have been christened as the 
Compact Symmetric Objects (CSOs, Wilkinson et al.\ 1994; Readhead et al. 1996a;  
Taylor, Readhead \& Pearson 1996; Owsianik, Conway \& Polatidis 1998)
which are believed to evolve into the 
Medium-sized Symmetric Objects (MSOs, Fanti et al.\ 1995)  and later to the 
standard large double-lobed (FR II) radio sources (e.g.\ Snellen 2003, and references therein)

High-resolution images of CSSs (e.g. Sanghera et al.\ 1995; Dallacasa et al. 
1995, 2002 and references therein) showed that 
there is a large range of structures associated with CSSs, ranging from double-lobed and
triple sources to those with complex structures.
Saikia et al.\ (1995, 2001, 2002) showed that,
as a class, the CSSs were more asymmetric than larger radio sources.
They also performed statistical tests of the orientation-based unified scheme, 
(Barthel 1989; Urry \& Padovani 1995), and 
concluded that there were differences between the CSS
radio galaxies (RGs) and quasars (QSRs) that were consistent with the
unified scheme; however, there was
also substantial evidence for more intrinsic asymmetries in CSSs than those which
were seen for the
larger sources.  Arshakian \& Longair (2000) studied the  FR II sources
in the 3CRR complete sample and also concluded that intrinsic/environmental
effects are more important for sources with small physical sizes; they also
found intrinsic asymmetries more important for sources of 
higher radio luminosities.
Furthermore, while comparing the hotspot-size ratio in 
radio sources, Jeyakumar \& Saikia (2000) found that CSSs are generally 
more asymmetric in this parameter as well; explaining such large
ratios of hotspot sizes seems to require an intrinsic origin for the size difference.

Detailed polarization observations of individual sources,
such as 3C147, show huge differential rotation measures between the two
oppositely directed lobes, suggesting their evolution in an asymmetric
environment (Junor et al.\ 1999). VLBA observations of 3C43 show evidence of high rotation
measure where the jet bends sharply, suggesting collision of the jet with a dense cloud of gas
(Cotton et al.\ 2003). 
The radio galaxy 3C459 has a highly asymmetric structure:
the eastern component is brighter, much closer to the nucleus
and is much less polarized than the western one, again suggesting
interaction of the jet with a dense cloud of gas. HST images of
the optical galaxy shows filamentary structures, suggestive of
tidal interactions (Thomasson, Saikia \& Muxlow 2003). Statistically,
CSS objects exhibit a higher degree of asymmetry in the polarization
of the outer lobes at cm wavelengths compared with the larger sources
(Saikia \& Gupta 2003; see Saikia et al.\ 2003 for a summary).
These asymmetries in the central regions of
active galaxies may be intimately related to the supply of fuel to the
central engine in these objects, possibly due to interactions with companion galaxies.

\begin{figure*}[t]
\begin{center}
 \psfig{file=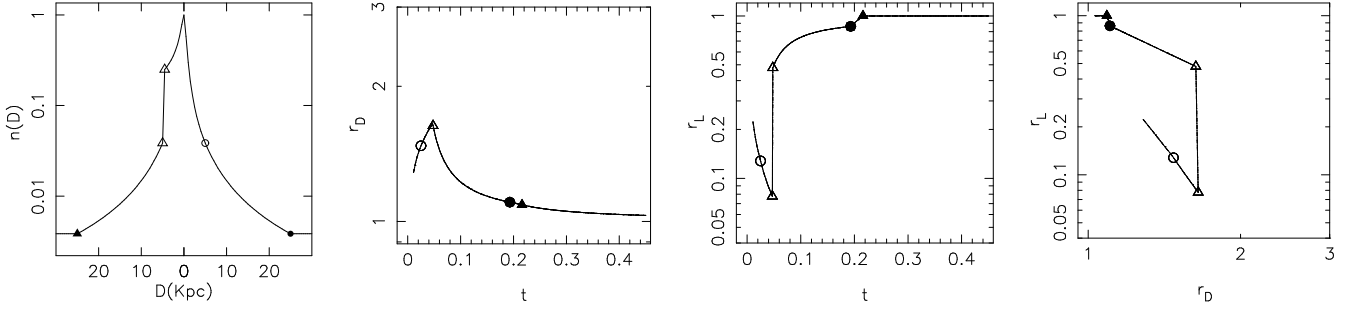,bbllx=24pt,bblly=580pt,bburx=587pt,bbury=711pt,width=17.8cm,clip=}
\end{center}
\caption[]{Analytic models for jets propagating through 
asymmetric media: (a) the density distributions, $n(D)$,
as a function of distance, with side 1 on the right and 2 on the left; 
since $\eta_1 = \eta_2$ there is a discontinuity
at $R_2$; (b) the arm-length ratio, 
$r_D(t)$; (c) the luminosity ratio, $r_L(t)$ and 
(d) the $r_D-r_L$ diagram, with both quantities now in log units.  
The open and filled circles
represent the times $T_1$ and $T_3$ respectively and the open 
and filled triangles represent the times $T_2$ and $T_4$ 
respectively. The unit of time is 1 Myr. \label{analytical}
}
\end{figure*}

All of these studies have indicated substantially greater asymmetries
for small radio sources. It is clearly worthwhile to attempt models of
their propagation so as to see if these differences can be understood. 
In this paper we develop a simple analytical model that enables us to treat
both arm-length asymmetries and luminosity asymmetries (Section 2). 
We then extend the analytical model to include relativistic effects so
as to allow properly for Doppler boosting and time delays if the
sources are ejected closer to the line-of-sight (Section 3). In Section 4 we 
summarize results of our two-dimensional and three-dimensional hydrodynamical 
models of jets propagating through
an asymmetric dense core, a symmetric galactic interstellar medium,
and finally, a symmetric uniform intracluster medium (ICM). We discuss the similar
results from the analytical and numerical models in Section 5.  

\section{Analytical Models for Jets Propagating in an Asymmetric Environment} 

To estimate the distances out to which jets with equal beam powers, but
propagating out through different interstellar media (ISMs), would reach at various times, we 
further generalize the ram pressure balance calculations discussed for
constant density media by Scheuer (1974) and extended to jets leaving a
power-law ISM and entering intracluster media by Gopal-Krishna \& Wiita (1987)
and Rosen \& Wiita (1988).  Such models assume that the jets start with 
conical cross-sections and that their heads, and associated hot-spots,
propagate sub-relativistically.  
We also estimate the ratios of radio luminosities of the two lobes
as functions of time by
expanding upon the calculations of Eilek \& Shore (1989) and Gopal-Krishna
\& Wiita (1991). The latter authors suggested that an ``excess'' ISM in the
nuclear regions (central few kpc), perhaps due to the interactions of galaxies, could
lead to a larger fraction of CSS sources at higher redshifts. Such 
interactions could both trigger a phase of jet production from the central
engine and easily lead to asymmetric distributions of gas in the inner
regions for at least the $<$10$^{7}$ yr  during which the jets propagate
through the galactic ISM.  The propagation through denser media also implies
higher efficiencies for conversion of jet power to radio luminosity (Gopal-Krishna
\& Wiita 1991; Blundell, Rawlings \& Willott 1999); this concept is supported 
by the recent findings of Arshakian \& Longair (2000) that more luminous 
sources are more asymmetric.

Our model calculations assume two intrinsically symmetric jets 
propagating outwards in opposite directions in pressure balance 
between the momentum flux density of the beam and ram pressure of the 
external medium, so the equation for the growth of one jet is 
(e.g., Scheuer 1974),
\begin{eqnarray}
 n(D){\left(\frac{dD}{dt}\right)}^2 &=& {4 K L_b} \over {\pi c \mu {m}_{h} {\left(D \theta \right)}^2}
\end{eqnarray}
where $K$ is an efficiency factor assumed to be a constant and of order unity, 
$\mu {m}_{h}$ is the mean particle mass where ${m}_{h}$ is the mass of the hydrogen atom,
$L_b$ is the power of the beam, $D$ is the distance from the nucleus and $\theta$ 
is the full-opening angle.  We assume the 
gas density of the interstellar medium (ISM) to be declining as
\begin{eqnarray}
n(D) = \eta n_0 \left[1+(D/a)^{2}\right]^{-\delta}. 
\end{eqnarray}
For $D \gg a$, $ n(D) \approx \eta n_0 { \left(D/a\right)}^{-2\delta}. $ 
This approximation is assumed for the analytical model considered here. To allow 
for asymmetrically distributed gas in the central regions, the slopes on 
opposite sides of the nucleus denoted by the subscripts 1 and 2,
$ \delta_{1,2} $ are taken as shallower, and the core radius $a_c$ smaller, 
in the interior regions, but the slope can be steeper in the
exterior regions. We have considered the possibility that the ISM 
density levels off to a constant ICM density at large radii. Standard central
densities of $n_0 \approx 10^{-1}$cm$^{-3}$ and $a \approx 2$ kpc are typical 
of large elliptical galaxies (Forman, Jones \& Tucker, 1985).  Excess central gas, 
requiring $\eta \approx 10$ and extending out to a few kpc (i.e., 
with $a_c \ll 1$ kpc) is sufficient to confine reasonably powered jets to 
CSS dimensions for significant lengths of time (e.g., Gopal-Krishna \& Wiita 1991;
Carvalho 1998).

We consider a model physically corresponding to jet propagation initially through
an asymmetric nuclear region, followed by a symmetrical galactic scale ISM, 
and finally, through a constant ICM.
The central ambient medium is characterized by 
$\eta=\eta_{1}$ and $\delta=\delta_{1}$ on one side, with the corresponding values
on the opposite side being $\eta_{2}$ and $\delta_{2}$. For the sake of 
simplicity, we take the core radius for both as
$a_{c}$.  The jet propagating outwards on side 1 reaches a distance $R_1$ after a time
$T_{1}$, while the one on the opposite side reaches $R_{2}$ 
 after a time
$T_{2}$.  Beyond $R_{1}$ and $R_{2}$, the density distribution on both sides is
described directly by equation (2), with $a \gg a_c$. 
The evolution after a distance $R_3 (> R_1 )$ on side 1 and
$R_4 ( > R_2 )$ on side 2, is governed by a constant 
density, with
$T_3$ and $T_4$ the times taken by  the corresponding jets to reach the
distances  $R_3$ and $R_4$, respectively (Fig.~\ref{analytical}).
Although we will usually assume $R_2 = R_1$ and $R_4 = R_3$, the times
at which those radii are reached will be different for the opposite
sides even for jets of
identical powers, and it is certainly possible that 
$R_2 \neq R_1$ or $R_4 \neq R_3$, so it is useful to retain this 
more general notation.  Quite clearly, the jet propagating through
the denser side (with shallower $\delta$ or higher $\eta$) will take
longer to get to any fixed distance.
Note that there must be a density discontinuity, either at $D = 0$ if
$\eta_1 \neq \eta_2$, or at $R_2$ if $\eta_1 = \eta_2$; the latter case is shown
in the left panel of Fig.~\ref{analytical}.
Such discontinuities
are most likely to arise through 
an asymmetric distribution of gas from a recent merger that may
be responsible for triggering the nuclear activity (e.g. Gopal-Krishna \& Wiita 1991).
However, it is relevant to note that any realistic situation would have steep
gradients and not discontinuities.
Note that even if $\delta > \delta_{1,2}$ (as explicitly assumed below),
because $a \gg a_c$ the 
absolute rate of decline in
density at $D>R_{1,2}$ is slower than at $D < R_{1,2}$ where the inner
core distributions dominate.

Using Eq.\ (2), for  $D \gg a$,
equation (1) reduces to
$D^{(1-\delta)}(dD/dt)=\eta^{-1/2}\tau^{-1}a^{2-\delta}$, 
where
\mbox{
$\tau={\left[{\mu n_0 c \pi {m}_{h} {\theta}^2} / {4 K L_b}\right]}^{1\over2}a^2.$}
Integrating the above equation yields 
$D = [ (2-\delta){\eta^{-{1\over2}} } 
{\tau^{-1}} t]^{1/(2-\delta)} a$.  
For $D=R$ at $t=T$, this can be expressed as $ D=R{({t/T})}^{1/(2-\delta)}.$
We use the above expression to estimate the distances
$D_1$ and $D_2$ to the lobes 1 and 2 respectively.

When the distances of propagation of the jet on both sides
satisfy $D_1<R_1$ and $D_2<R_2(=R_1)$
then 
\begin{eqnarray}
 D_{1,2}=R_{1,2} {({t/T_{1,2}})}^{1/(2-\delta_{1,2})}. 
\end{eqnarray}
But once 
$D_1>R_1$,
\begin{eqnarray}
D_1 = {\left[{{{2-\delta} \over {2 - \delta_1}} \left( {{t \over T_1} -1} \right) 
+ 1 }\right] }^{ 1/(2 - \delta)} R_1.
\end{eqnarray}
For $t\gg T_1$ this can be expressed as \\
\mbox
{$D_1=  R_1 \left([(2-\delta) / (2-\delta_1)][t/T_1]\right)^{1/ (2-\delta)}$.}
Similarly, for the jet propagating on the opposite side, 
for $D_2>R_2$,
\begin{eqnarray}
D_2 = {\left[  { F
\left( {{t \over T_2} -1} \right) + 1 } \right] }^{ 1/(2 - \delta)} R_2, \ \
\end{eqnarray} 
\noindent with
\begin{eqnarray*}
F = {{2-\delta} \over {2 - \delta_2}} \left({{\eta_2} \over {\eta_1}}\right)^{1\over 2}
{(a_c/R_2)^{\delta_2-\delta} \over (a_c/R_1)^{\delta_1-\delta}}
= {{2-\delta} \over {2 - \delta_2}}\left({{a_c} \over {R_1}}\right)^{\delta_2-\delta_1},
\end{eqnarray*} 
\noindent where the last equality holds when $\eta_1 = \eta_2$ and
$R_1 = R_2$.  For $t\gg T_2$, 
$D_2= R_2(F/T_2)^{1/(2-\delta)} t^{1/(2 -\delta)}$.

We now estimate the ratios of separations, $r_D$, and luminosities, $r_L$, for
the lobes during different stages in the evolution of the source for this model. 
We choose the 
slopes of the density distributions such that  $\delta > \delta_1 > \delta_2$, and 
$ \eta_2 = \eta_1$
so that  $D_1$ is greater than or equal to $D_2$ when evaluated at the same time,
though we leave some dependences on $\eta$ explicit in the following expressions.
For $t < T_1$ and $t < T_2$, when both the jets are within the asymmetric
environment, 
$r_D= [T_2^{1/(2 - \delta_2)}/ T_1^{1/(2-\delta_1)}] t^{(\delta_1 - \delta_2)/[(2 -\delta_1)(2-\delta_2)]}$.
Using the approximate result that the radio luminosity on each side scales as
$L_R \propto n(D)^{3(1+\alpha)/10} \propto n(D)^m $ (Eilek \& Shore 1989;
Gopal-Krishna \& Wiita 1991) we find that
\begin{eqnarray*} 
r_L = \left[{\eta_1 \over \eta_2} \left({R_1 \over a_c}\right)^{-2(\delta_1-\delta_2)} 
T_1^{2\delta_1 \over 2-\delta_1} 
T_2^{-2\delta_2 \over 2-\delta_2} t^{-4(\delta_1-\delta_2) \over (2-\delta_1)(2-\delta_2)} \right]^m,
\end{eqnarray*}
where we have taken $R_1 = R_2$.
With $\alpha \approx 1.0, m \approx 0.6$. For the case shown in Fig.\ 1, where
$\delta=0.75$, $\delta_1=0.675$ and $\delta_2=0.3$, $r_D \propto t^{0.17}$, and
$r_L \propto t^{-0.40}$. Thus, at this early phase of the evolution, 
$r_D$ increases with time while $r_L$ decreases with time.

For $t \gg T_1$ and $t < T_2$,
\begin{eqnarray*} 
r_D = \left({2-\delta \over 2-\delta_1}\right)^{1 \over 2-\delta}  T_1^{-1 \over 2-\delta} 
T_2^{1 \over 2-\delta_2} 
t^{ \delta - \delta_2 \over  (2 -\delta)(2-\delta_2)};
\end{eqnarray*}
\begin{eqnarray*}
r_L = \left[{\eta_1 \over \eta_2}\left({2-\delta \over 2-\delta_1}\right)^{-2\delta \over 2-\delta}
\left({R_1 \over a_c}\right)^{2(\delta_2-\delta_1)} 
{T_1^{2\delta \over 2-\delta} \over T_2^{2\delta_2 \over 2-\delta_2} }
t^{-4(\delta-\delta_2) \over (2-\delta)(2-\delta_2)} \right]^m,
\end{eqnarray*}
\noindent where we have used the continuity of density at $R_1$ to eliminate
$\eta$ and $a$ through the relation
$\eta a^{2\delta}  = \eta_1 a_c^{2\delta_1} R_1^{-2 (\delta_1 - \delta) }$. 
For the above values of the $\delta$'s and $m$ we see that the rise in $r_D$ and
the decline in $r_L$ both become slightly stronger, with $r_D \propto t^{0.21}$
and $r_L \propto t^{-0.51}$.

For $t > T_3$, when the faster jet crosses $R_3$, the distance travelled by lobe 
1 is given by 
\begin{eqnarray}
D_1 = {\left[{2 \over {2 -\delta_1}} \left({R_1 \over R_3}\right)^{2-\delta}  
        \left({t\over T_1} - {T_3 \over T_1} \right) +1 \right]}^{1\over2} R_3.
\end{eqnarray}
\noindent For $t \gg T_3$ this result can be expressed as 
\mbox{
$ D_1 = R_3 [2/(2-\delta_1)]^{1\over 2}
(R_1/R_3)^{1-\delta/2} (t/T_1)^{1/2}$. }

During the last intermediate phase, when $t \gg T_3$ but $T_2 \ll t < T_4$,
\begin{eqnarray*}   
r_D = \left({2 \over 2-\delta_1}\right)^{1/2} F^{-{1 \over 2-\delta}}  
\left({R_3 \over R_1}\right)^{\delta \over 2}
T_1^{-{1 \over 2}} T_2^{{1 \over 2-\delta}} t^{-{\delta \over 2(2 -\delta)}},
\end{eqnarray*} 
\noindent and 
\begin{eqnarray*}
r_L = \left[ \left( {R_1 \over R_3} \right)^{2\delta} \left( {F t \over T_2} \right)
^{2\delta/(2-\delta)} \right]^m.
\end{eqnarray*}
For the above choice of the parameters for the density profile, $r_D \propto t^{-0.30}$. Thus
at moderately late times $r_D$ slowly decreases with time, while $r_L$ 
more rapidly approaches unity from below, since $r_L \propto t^{0.72}$. 

Finally, once $t > T_4$, and the slower progressing lobe also enters the uniform
intracluster medium,
\begin{eqnarray}
D_2 = \left[ G \left({t\over T_2} - {T_4 \over T_2}\right) +1 \right]^{1\over2} R_3, 
\end{eqnarray}
\noindent where

\begin{eqnarray*}
G = {\left[{2 \over 2 -\delta_2} \left({\eta_2 \over \eta_1}\right)^{1\over 2}
a_c^{\delta_2-\delta_1}R_1^{\delta_1-\delta_2+2-\delta}R_3^{\delta-2}\right]}.
\end{eqnarray*}
For $t \gg T_4$, this simplifies to  $ D_2 = R_3 G^{1 \over 2} (t/T_2)^{1 \over 2}$.
Thus, when the jets have traversed well beyond $R_3$ and $R_4$, so that $t \gg T_3, T_4$,  
$r_D  \longrightarrow 1 $,  while
$r_L \simeq 1$ as soon as both lobes are in the constant density ICM. 

In summary, the size ratio grows monotonically until $T_2$, after which it
declines slowly toward unity.  The flux-density ratio varies more dramatically, and
with more complexity,
dropping quite a bit until $T_2$ and  then rising rapidly at that time, because of our
assumption of a density discontinuity at $R_2$.
Both before and after $T_3$, the flux-density ratio of the components 
 becomes symmetric
more rapidly than does the separation ratio,
 with $r_L \propto r_D^{-4m}$,
but because of its earlier severe deviations from unity, it remains 
more asymmetric.
 Finally, once lobe 2 also crosses the outer interface (at  $T_4$) and the constant density
ICM is entered, the ratio $r_L$ very quickly approaches
unity, but $r_D$ does so only asymptotically.

\subsection{Results of basic analytical models}  
Fig.~\ref{analytical} shows the variation of density with distance from
the nucleus on opposite sides,  the variations of 
the separation ratio, $r_D$, and the flux density ratio, $r_L$, 
with time as the jets propagate outwards, as well as an $r_D$--$r_L$ diagram.
The time axis is in units of $10^6$ yr.
The times $T_1$, $T_2$, $T_3$, and $T_4$ 
when the jet crosses the interfaces at $R_1$, $R_2$, $R_3$, and $R_4$, 
respectively, are marked on the trajectories.
Our calculations of such trajectories provide an explanation for the observed
points in the $r_D-r_L$ diagram which have $r_D$ greater than 1 and $r_L$ 
smaller than 1 (see Saikia et al.  1995, 2001).
Our calculations also show that after the jets have left the 
asymmetric environment, and are passing through similar media on opposite
sides, the source tends to become symmetric quite rapidly. Thus most
large FR II sources, which tend to be quite symmetric in both distances 
(cf.\ Arshakian \& Longair 2000) and
luminosities of the oppositely directed lobes, may still have passed through an 
asymmetric environment in early phases of their evolutions. This is consistent
with the possibility that most sources may have undergone  CSS phases during
the early stages of their evolutions, as discussed in Section 1. 
In only a few of the CSS sources, such as
3C 119 (Ren-dong et al.\ 1991) and  3C 48 (Wilkinson et al.\ 1991), 
does it appear that the jets may have been disrupted by interaction 
with the clumpy external medium so that they never evolve into  large sources
(e.g.\ De Young 1993; Higgins, O'Brien, \& Dunlop 1999; Wang, Wiita, \& Hooda 2000).

For a specific reasonable example, 
consider a jet of a beam power of $7\times10^{46}$
erg s$^{-1}$ and an opening angle of 0.04 rad,
 propagating through a medium with $n_0=1$ cm$^{-3}$,
core radii $a_c = 0.45$ kpc and $a = 2.0$ kpc, with $\delta$=0.75, $\delta_1$=0.675, $\delta_2$=0.3.
 The times required to reach $R_1$=5 kpc, 
$R_2$=5 kpc, $R_3$=25 kpc, and $R_4$=25 kpc are 0.025, 0.047, 0.19, and 0.22
$\times$ 10$^{6}$ yr, respectively. The above times for a weaker jet of 
power $2\times10^{45}$ erg s$^{-1}$, with an opening angle of 0.1 radian propagating 
in a medium with smaller $a_c = 50$ pc and higher 
central density $n_0=10$ cm$^{-3}$, but otherwise identical parameters,
are 0.26, 1.1, 2.0 and 2.9 $\times$ 10$^{6}$ yr, respectively. These ages are
smaller than any plausible age estimates for larger sources (e.g.\ Leahy, Muxlow \&
Stephens 1989; Scheuer 1995; Blundell \& Rawlings 2000).
The brief times during which our models produce sources which
would be classified as GPSs (well within $R_1$) are consistent with the different analytical
approximations employed by Carvalho (1998) for jets passing through
extra dense (but uniform) matter.

If there is substantially less gas on one side than the other, as is the case
either if the power-law of the central fall-off is substantially less steep on 
one side than on the other, say $\delta_{1} - \delta_{2}\approx 0.4$,
or if $\eta_2 \approx 4 \eta_1$,
we can obtain fairly wide ranges in the predicted ratios of separations and 
radio luminosities, and explain the very asymmetric sources. For most sources,
a modest difference in $\delta$ of about 0.15 turns out to be adequate to
produce the observed asymmetries, even assuming symmetric 
jets of equal powers are launched from the central engine.

\section{Relativistic and Viewing Angle Effects} 

For more detailed comparison with the observations, 
two other important effects must be included:
the observed luminosities are boosted or depressed via possible bulk relativistic
motions; and the observed positions are affected by light travel time effects.

In this section the subscript `a' will denote the lobe approaching the observer
and the subscript `r' will denote the receding component. So
 $D_a$ and $D_r$ are the distances from the core to the approaching and receding lobes,
respectively, while
$D_{1,2}(t)$ are the distances traveled 
by the jets propagating through the less dense and denser media, 
respectively, as in the previous section;
let $\varphi$ be the angle between the approaching jet and the line of sight to
the observer. 

In the core's rest frame we are observing at time $t$.  The corresponding times at which
emission arises from the approaching/receding lobe are:
$t_a = t/(1 - \beta_a {\rm cos}\varphi)$; 
$t_r = t/(1 + \beta_r {\rm cos}\varphi)$, where $\beta = v/c$.
Including light travel-time effects, where the appropriate velocity is 
the mean speed of advance to that point, 
the {\it observed} distances of the approaching and receding lobes are:
\begin{eqnarray*}
d_a(t) = D_a(t) {\rm sin}{\varphi}/(1 - [D_a(t)/(ct)] {\rm cos}\varphi);  \\
d_r(t) = D_r(t) {\rm sin}{\varphi}/(1 + [D_r(t)/(ct)] {\rm cos}\varphi).
\end{eqnarray*}
Thus, the distance ratio, including projection and light-travel time effects is
\begin{eqnarray}
r_D^*(t) \equiv {\frac{d_a(t)}{d_r(t)}} =\frac{D_a(t)}{D_r(t)} ~~{\frac{1 + [D_r(t)/(ct)] {\rm cos}\varphi}
{1 - [D_a(t)/(ct)] {\rm cos}\varphi}},
\end{eqnarray}
where the superscript $*$ indicates that relativistic effects have been
included.

The luminosity corrections also will depend upon the Doppler boosting
factor, which could be significant if much of the emission comes from the hotspots
moving at mildly relativistic speeds, which is plausible for young sources,
such as GPS and CSS objects. 
For a hot-spot or knot within a jet, bulk motions produce a change in 
flux density, so that (Scheuer \& Readhead 1979)
\begin{eqnarray*}  
S_{obs (a,r)}(\nu) = S_{em}(\nu)[\Gamma (1 \mp \beta {\rm cos}\varphi)]^{-(3+\alpha)},
\end{eqnarray*}
where the Lorentz factor, $\Gamma \equiv (1-\beta^2)^{-1/2}$. 
The suitable values of $\beta$ are 
the instantaneous rates of advance of the lobes. 

In the notation of \S 2, $S_{em} = L_R$ and $L_R \propto n(D)^m$.
Therefore, the above equations tell us that the observed luminosity
ratio is:
\begin{eqnarray}
r_L^*(t) = {\frac{S_{obs,a}(t)}{S_{obs,r}(t)}} 
 = {\frac{L_{R,a}(t_a)[\Gamma_a^*
(1-\beta_a^*{\rm cos}\varphi)]^{-(3+\alpha)}}
{L_{R,r}(t_r)[\Gamma_r^*(1+\beta_r^*{\rm cos}\varphi)]^{-(3+\alpha)}}}.
\end{eqnarray}
Where
$\beta_a^* \equiv {\frac{1}{c}}{\frac{dD_a(t_a)}{dt_a}}$, ~
$\beta_r^* \equiv {\frac{1}{c}}{\frac{dD_r(t_r)}{dt_r}}$, and
$\Gamma^*_{a,r}$ are determined from these
 {\it instantaneous} $\beta^*$s evaluated at the 
approaching (or receding) times.
To evaluate this precisely, albeit in a model dependent fashion,
explicit formulae for $L_{R,a}(t_a)$ and
$L_{R,r}(t_r)$ should be used (e.g., Eilek \& Shore 1989; Blundell, Rawlings \& Willott 1999; 
Manolakou \& Kirk 2002), and no further simplification is possible.
However, the boosting term will typically dominate over 
the light-travel time factors for the luminosity ratio, even for the low values of $\beta \le 0.2$
derived from spectral aging and other considerations (Scheuer 1995; Gopal-Krishna
\& Wiita 1996; Arshakian \& Longair 2000).  Therefore, for analytical work
Eq.\ (9) can be approximated as
\begin{eqnarray}
r_L^*(t) \simeq  {\frac{L_{R,a}(t)}{L_{R,r}(t)}}~ {\frac {[\Gamma_a^*
(1-\beta_a^*{\rm cos}\varphi)]^{-(3+\alpha)}}
{[\Gamma_r^*(1+\beta_r^*{\rm cos}\varphi)]^{-(3+\alpha)}}}. ~~~~~~~~~~
\end{eqnarray}

\begin{figure*}[ht]
\hbox{
\includegraphics[width=15.5cm,bbllx=12pt,bblly=365pt,bburx=596pt,bbury=711pt,clip=]{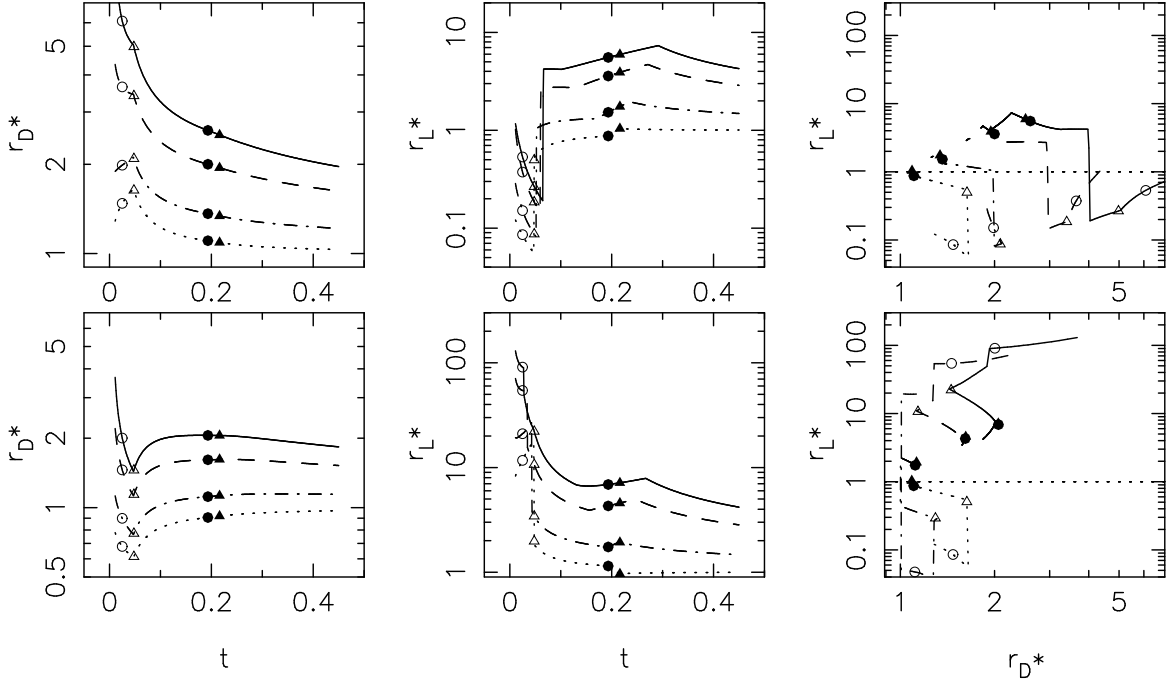}
}
\caption{
Evolutionary tracks for $r_D^*, r_L^*$ and  $r_D^*-r_L^*$ diagrams for different viewing angles to the line of sight. The upper panels are for Case 1 while the lower  panels are for Case 2. The solid, dashed,  dot-dashed and dotted lines indicate a viewing angle, 
$\varphi$, of 10, 45, 75 and 90 degrees, respectively.
The symbols marking the time have same meaning as in 
Fig.~\ref{analytical}, and the unit of time is 1 Myr. 
\label{fig2}
}
\end{figure*}

\subsection{Evolutionary tracks in the $r_D-r_L$ diagram}
{\bf Case 1:}
Here we consider the situation where the jet approaching us is propagating 
through the less dense medium.
Then both density and travel-time
effects act to increase the observed arm-length ratio, $r^*_D$.
However the  travel-time and density induced luminosity changes work against
 the relativistic effects in finding the luminosity ratio, $r^*_L$. 
Since the side approaching us comes through a lower density medium, it
is both farther out and observed at a ``later'' time, so that the density
is even lower (or at most the same) as it was for the receding lobe; both of these
effects imply a lower emissivity; however, the approaching lobe is Doppler boosted.
Meanwhile, the receding lobe is still in a higher density region (at 
least until both have reached the ICM), and is observed at
an earlier time in its history, so $n(D_r)$ is even higher, implying relatively
enhanced emissivity; but it is weakened by the Doppler effect.

Here the appropriate values are $D_a(t)=D_1(t)$ and $D_r(t)=D_2(t)$, where
$D_{1,2}$ are obtained 
from Eqs.\ (3--7), with the appropriate equations used depending
upon the distances the two lobes have travelled. 
Then, recalling
that $r_D(t) \equiv D_1(t)/D_2(t)$ we have for the observed ratio,
\begin{eqnarray}
r_D^*(t) \equiv {\frac{d_a(t)}{d_r(t)}} = r_D(t) ~~\left[ {\frac{1 + [D_2(t)/(ct)] {\rm cos}\varphi}
{1 - [D_1(t)/(ct)] {\rm cos}\varphi}} \right].
\end{eqnarray}
and
\begin{eqnarray}
r_L^*(t) \simeq  r_L(t) ~~\left[{\frac {\Gamma_2^*
(1+\beta_2^*{\rm cos}\varphi)}{\Gamma_1^*(1-\beta_1^*{\rm cos}\varphi)}}\right]^{(3+\alpha)} 
\equiv ~r_L(t) {\cal D}_1(t).
\end{eqnarray}

The observed $r^*_D$ is increased more for smaller viewing angles as shown
in the upper-left panel of Fig.\ 2. 
Although $r_L$ can be quite small at very early times,
the ratio of the Doppler boosting factors, ${\cal D}_1$, is usually
greater than unity, and this raises  $r^*_L$.  Even for modest $\beta (< 0.2)$
it can invert the ratio and let the lobe more distant from the
nucleus  appear brighter {\it
if it is approaching} the observer. 
At early  times, the density effects dominate and 
the sources fall below the $r^*_L=1$ line, unless $\varphi$
is very small, in which case $r^*_L$ may initially exceed unity
before quickly dropping below it. Later on, the value of $r_L$
rises above 1 for $\varphi < 90^{\circ}$, and for most viewing
angles
 the source spends a considerable amount of time within a factor of a few 
of the $r_L=1$ line, as shown in the upper-middle panel of Fig.~\ref{fig2}.
The evolutionary tracks move to the right on the $r_L^* - r_D^*$ plane as the
jet points closer to the line-of-sight (Fig.\ 2, upper-right). 
Therefore, this scenario can explain  sources observed to have
fairly substantial ($\sim 10$) asymmetries in distance and/or luminosity; those
asymmetries can be correlated or anti-correlated, depending
on viewing angle and evolutionary phase.

\noindent{\bf Case 2:} In this situation the side approaching us comes through the 
denser medium, so $D_a(t) = D_2(t)$ and $D_r(t) = D_1(t)$.  Now,
\begin{eqnarray}
r_D^*(t) \equiv {\frac{d_a(t)}{d_r(t)}} = {\frac{1}{r_D(t)}}~~ 
{\frac{1 + [D_1(t)/(ct)] {\rm cos}\varphi}
{1 - [D_2(t)/(ct)] {\rm cos}\varphi}},
\end{eqnarray}
and
\begin{eqnarray}
r_L^*(t)  \simeq \frac{1}{r_L(t)}  {\left[\frac{\Gamma_1^* (1+\beta_1^* {\rm cos}\varphi)}
{\Gamma_2^* (1-\beta_2^* {\rm cos}\varphi)}\right]}^{3+\alpha}
\equiv  \frac{{\cal D}_2(t)} {r_L(t)}.
\end{eqnarray}

Since $1/r_L(t) > 1$, both of the factors in the above 
expression enhance the luminosity 
ratio, $r_L^*$, and at very early times it can be very high. 
However, the two effects work oppositely for the arm-length ratio, $r_D^*$. 
Although $1/r_D(t) < 1$, for smaller viewing angles and  at very early times, 
$r_D^*$ can be well above unity; such sources have $r_L^* > 1$ 
but are less asymmetric in $r_D^*$ compared to Case 1,
as shown in Fig.~\ref{fig2}. 
Up until $T_2$, $r_D^*$ falls; around then the density effects 
are comparable to the 
light travel time effects.  At later times $r_D^*$ rises a bit,
but by $T_4$ it starts to asymptotically approach unity.
Although we have plotted $r_D^*$ (from Eq.\ 13) as less than 1 in the lower-left
panel of Fig.\ 2,  since the observed arm-length ratio is defined to be $>1$, 
we invert both $r_D^*$ and $r_L^*$ 
for the $\varphi = 90^{\circ}$ curve in the bottom-right panel of Fig. 2,
showing that the evolutionary track in the $r_D^* - r_L^*$ plane
is similar to that of an unbeamed source (Fig.\ 1). 
A similar procedure is adopted for $\varphi = 75^{\circ}$ in the time range
when $r_D^* < 1$.
The typical outcome of this  scenario is a source observed to be 
nearly symmetric in $r_D$ but possibly very asymmetric in $r_L$. 

\begin{figure*}
\psfig{file=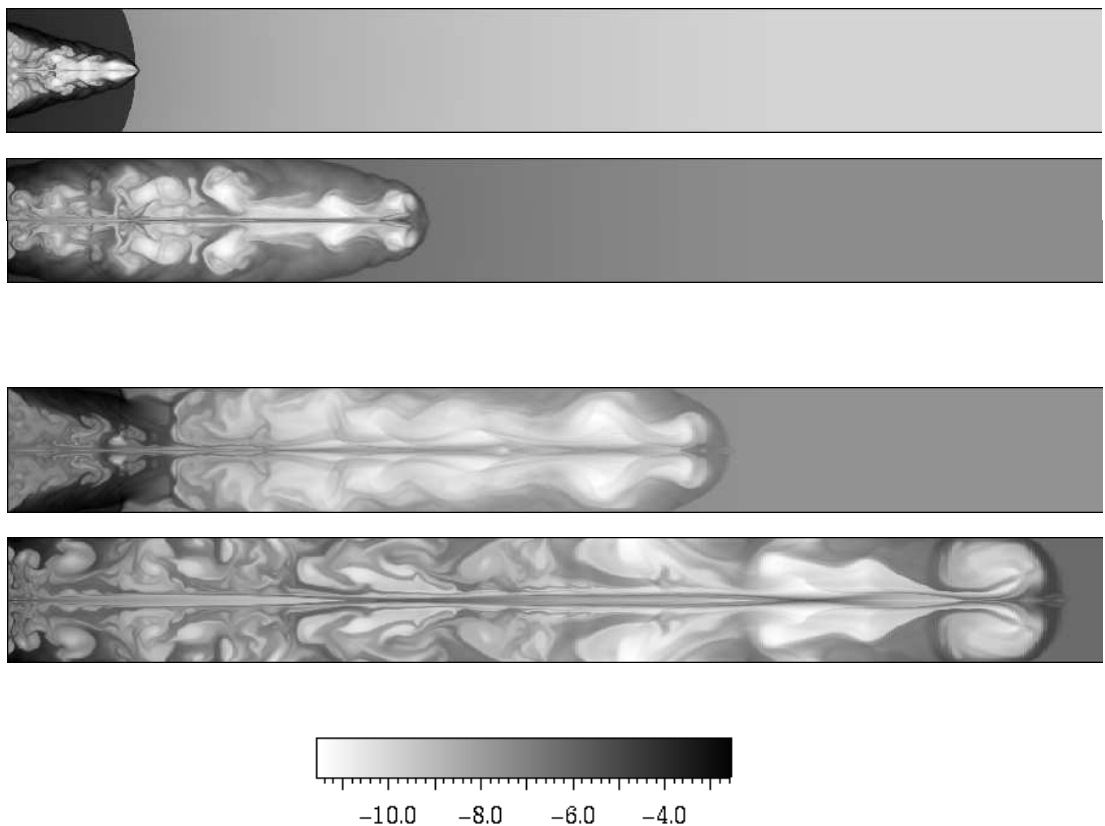,bbllx=56pt,bblly=235pt,bburx=572pt,bbury=637pt,width=6.0in}
\caption{Two-dimensional simulations, set up to nearly
reproduce our analytical model in the plane of the sky
($\varphi = 90^{\circ}$).  The natural logarithm
of the density
is shown for Runs A2 (upper image in each panel) and B2
(lower) at $t=0.80$ (top panel)
and $t=2.05$ (bottom panel). The computation is conducted
on the upper
half of the illustrated region and symmetry about the
propagation axis
is assumed in plotting the density.  The total length of the grid is 35 kpc,
with $R_1=$ 5 kpc and $R_3 =$ 25 kpc; here the unit of time is 0.6 Myr.
\label{2Dimage}
}
\end{figure*}

\section{Numerical Simulations} 
We have examined some of our analytical estimates using numerical
simulations of propagating jets. To follow the jets with very high
resolution for the extended 
distances that are necessary if we are to examine their evolution 
for situations involving the multiple external media in our models,
we are constrained to perform two-dimensional axisymmetric computations. 
To allow for some physically realistic breaking of axisymmetry, we have also
performed three-dimensional simulations, albeit at lower resolution.
We used the well-tested general purpose ZEUS 2D  and 3D (v.3.4) codes which are 
described in depth in Stone \& Norman (1992a,b), Stone, Mihalas, \& Norman (1992), 
Clarke \& Norman (1994), and Clarke (1996a).  
As we could not implement the magnetohydrodynamical
version of ZEUS (e.g., Clarke 1996b), we cannot make even the quasi-realistic
estimates of the source luminosities that can arise from codes that 
include magnetic fields.  Therefore, these simulations are useful for
examining $D(t)$ and $r_D(t)$, but not $L(t)$ or $r_L(t)$.
We now briefly summarize results from these efforts.

\subsection{2-D Simulations} 
For the first set of simulations we used the ZEUS 2-D code 
as modified by Hooda et al.\ (1994). 
Our 2-D work used cylindrical coordinates with 900 zones
along the axial (jet propagation) direction and 200 zones in the radial
direction.  The jet radius has been scaled to 0.05 kpc, with the
entire grid running from 1--35 kpc in the axial direction and 0--2 kpc in the radial
direction.  With these choices, 25 zones are included within the initial jet radius, and
adequate resolution of internal shocks and rarefactions are achieved.

The density within the ISM exactly follows the profile as defined by Eqn.\ (2),
and does not make the pure power-law assumption used in \S 2.
In this work the important parameters are:  the jet Mach number, 
${M_j = {u}/{a_{\rm in}}}$, with $a_{\rm in}$ the jet sound speed and $u$ the
velocity of the jet;
$\zeta \equiv \rho_{\rm jet}$/$\rho_0$, with $\rho_0$ the initial ambient
density and $\rho_j$ the density of the jet. Our numerical simulations 
also assume initially conical jets, i.e., 
we set the half opening angle to 0.05 rad in these computations 
(cf.\ Wiita, Rosen \& Norman 1990; Wiita \& Norman 1992.) 

\begin{table}[h]
\caption{The parameters of the numerical runs}
\label{ptable}
\begin{tabular}{l|ll|ll}
\hline
 & \multicolumn{2}{c|}{2D}  & \multicolumn{2}{c}{3D} \\
 & A2 & B2 & A3 & B3 \\
\hline
$\zeta$  & 0.001  & 0.001 & 0.001 & 0.001 \\
$M_j$   & 26  & 26 & 26  & 26 \\
$\theta_j$(radian)   & 0.05  & 0.05 & 0.02  & 0.02 \\
$R_1$(kpc)   & 5  & 5 & 5 & 5 \\
$R_3$(kpc)   & 25  & 25 & 25 & 25 \\
$a_c$(kpc)   & 0.05  & 0.05 & 0.45 & 0.45 \\
$a$(kpc)   & 1.0  & 1.0 & 2.0  & 2.0 \\
$\delta$   & 0.75  & 0.75 & 0.75 & 0.75\\
$\delta_1$   & $-$  & 0.675  & $-$ &  0.675\\
$\delta_2$   & 0.3  & $-$ & 0.3  & $-$\\
$R_0$(kpc)   &0.05  &0.05 & 1.0 & 1.0 \\
\hline
\end{tabular}
\end{table}

We performed two simulations (A2, B2, where A2 corresponds to $D_2$
and B2 to $D_1$ of Sect. 2) with 
parameters given in Table 1.
Beyond 25 kpc both densities were
constant and equal. 
In both cases, the jets are assumed to be in initial pressure balance 
with the external medium. All individual portions of the external gases 
are taken as isothermal 
and the interfaces are pressure matched; thus in Case A2 there is an
abrupt density drop by a factor of 0.03 at the first interface at 5 kpc
(easily seen in the uppermost panel of Fig.~\ref{2Dimage}), and there is
a corresponding rise in $T$ by a factor of 33 across that interface. 

\begin{figure}[t]
\psfig{file=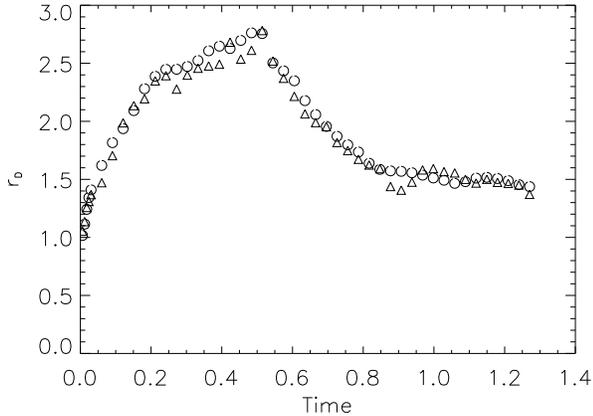,width=3.5in}
\caption{The ratios of the location of the Mach disk and 
bow shock
of Run B2 to those of A2, $r_D$, as  functions of time 
scaled to units
of Myr. The triangle denotes the bow shock $r_D$ and the 
circle denotes the Mach disk
$r_D$.  
}
\end{figure}

Although the code is not relativistic, the simulation can be 
reasonably scaled so that it is applicable 
to FR II radio sources. For a value of the initial jet radius $R_0$ = 0.05 kpc 
and  $T_{ISM}$ = 10$^6$K, one time unit corresponds to 0.6 $\times$ 10$^6$yr.
The velocity of the jet flow with respect to the ambient medium is $0.4c$ 
and the power of the jet is $L_b \simeq 2\times10^{45}$  erg s$^{-1}$ for a central
ambient density of $n_0 = 10$ cm$^{-3}$. 

Figure~\ref{2Dimage} shows gas density for the two 
simulations (A2 and B2) at the same early (set of two at the top) and late 
(2.5 times greater; set of two at the bottom) times. 
The early time is chosen to correspond to the point when the slower 
jet (A2) just emerges from the denser inner core. At that same time, the identical 
jet propagating through the less dense inner core (B2) has long since blasted 
though its more rapidly declining inner medium 
($\delta_{B2} \equiv \delta_1 > \delta_{A2} \equiv \delta_2$), 
and has gone a good way out into the symmetric ISM.  So at that time, the ratio 
of distances propagated is nearly maximal, since the 
asymmetry produced by the differences between the central media is
fully felt.  At the late time shown, just before the less impeded jet leaves the grid, 
the ratio of distances is substantially reduced, in that the early asymmetry in
environment is mitigated by the longer phase of propagation through symmetric
media.  
Figure 4 illustrates the 2-D results for $r_D$ defined using both
the Mach disk and the bow shock positions; we see that for these simulations
there is essentially no difference between them, as the jet always terminates
just a short distance behind the bow shock.
Since CSS sources are restricted by definition to sizes in which 
galactic halos and any irregularities in them will play a large role, the
much greater asymmetries observed for them (as compared to substantially bigger FR II
RGs) are easily accommodated.  

\begin{figure*}[t]
\psfig{file=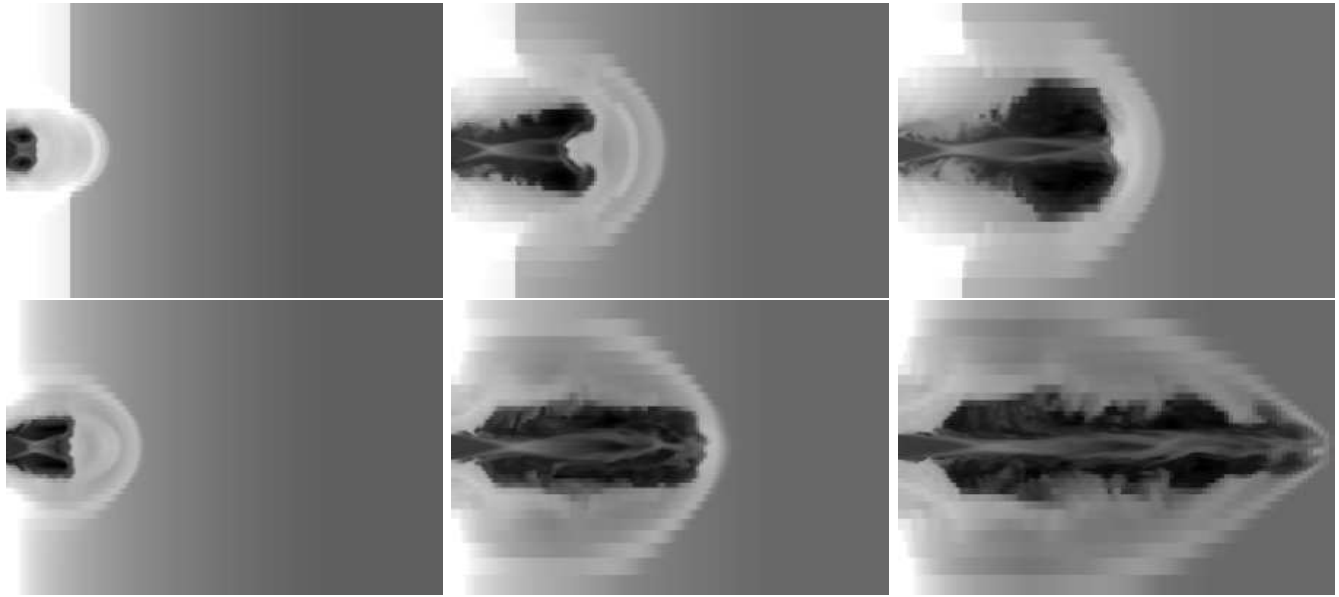,bbllx=80pt,bblly=498pt,bburx=545pt,bbury=709pt,width=7.0in}
\caption{The natural logarithm of the density (with white the densest and black the
least dense) is shown for
Runs A3 (top) and B3 (bottom) at three times:
$t= 0.021$, left; $t=0.080$, middle; $t=0.144$, right; here the unit of time
is 6.0 Myr.
Note that these are slices through a 3-D computation and no
symmetries are imposed, so that non-axisymmetric motions are clearly visible in the rightmost
 panels. The grid runs from 1--35 kpc along the propagation direction and $\pm 12$kpc
in the transverse direction.
\label{3Dimage} }
\end{figure*}

\subsection{3-D Simulations} 
Three-dimensional simulations allow for the generation of 
non-axisymmetric modes and instabilities.
The parameters of our 3-D simulations are 
also given in Table 1, and are similar to those of the 
2-D simulations. 
The main differences are the need to use a larger values of  $R_0$ (1.0 kpc),
$a_c$, and $a$ and a somewhat smaller value of $\theta_j$,
these changes being necessary to allow sufficient resolution in the necessarily
coarser 3-D simulations.
These three-dimensional simulations were run on 
150$\times$50$\times$50 active zones,
extending out to 35$R_0$ (35 kpc) along the initial direction of propagation, 
$x$, and  to $\pm$12$R_0$ ($\pm$12 kpc) in both directions perpendicular to it. 
To provide marginally adequate resolution in the most important region, 14 
uniform zones span the jet diameter, with the rest of the zones
logarithmically scaled (e.g. Hooda \& Wiita 1996; 1998). 
This set-up allows the jets sufficient time to propagate through each medium 
so that the effects of the different ambient media on the jets 
can be ascertained; see Saikia et al.\ (2003).  We believe these to be the first 3-D extragalactic jet
simulations to employ non-zero opening angles. 

With the exception of  a relatively wider jet
(in comparison to the size of the core and main galaxy, which is required to
provide adequate resolution within the jet), and minor differences
in the scale heights, the parameters for the
first two 3-D simulations are identical to those of the 2-D
simulations. Therefore we can quite quickly see the differences produced
by adding a third dimension. 

Figure~\ref{3Dimage} displays densities from simulations A3 and B3 at early, middle and 
late times in the evolutions.
For a canonical value of $R_0$ = 1 kpc and $T_{\rm ISM} = 10^6$K,
the unit of time is now 6.0 $\times$ 10$^6$ yr. 
A major difference between the 2-D and 3-D runs is that the bow shocks run ahead
of the Mach disks by much greater amounts in 3-D than they do in 2-D.
At early times, when the jet is still in a transient initial state, the bow shocks for these $M_j = 26$ jets
proceed out through the inner halo very quickly, and open a large gap between
the bow shocks and the working surfaces at the ends of the jets. 
Our analytical models assume that these two locations stay very close together, 
so these simulations reveal that this assumption is an oversimplification 
for such high Mach numbers and low jet densities.
A similar behaviour is also seen in axisymmetric simulations, using a different
3-D code, of very light cylindrical
jets propagating through constant density media (Krause 2003).

 The finite opening angles
are fairly well preserved in the first phases of the evolution (both in the central
halo and for some distance into the more steeply declining ISM), despite the
presence of a recollimation shock.
Once the jets are at a substantial distance into the ISM, the jet becomes
better collimated and then propagates nearly as a cylinder (see the middle panel
of Fig.~\ref{3Dimage} for B3 and the right panel for A3).  At this point,
our analytical models, which assume the opening angle stays constant,
lose more of their validity, as the slowing down of propagation predicted by
the spreading jet picture is replaced by a situation of more nearly constant
velocity. 
We see a slow breaking of axisymmetry through these simulations, despite
the symmetric initial conditions.  This can be attributed primarily to numerical
errors introduced by our setting up of a conical flow within a relatively coarse 
Cartesian grid.  Unsurprisingly, the non-axisymmetric effects
are stronger for A3, which crosses a density jump; such a discontinuity 
helps to induce the growth of perturbations (cf.\ Hooda \& Wiita 1996, 1998).

\begin{figure}[t]
\psfig{file=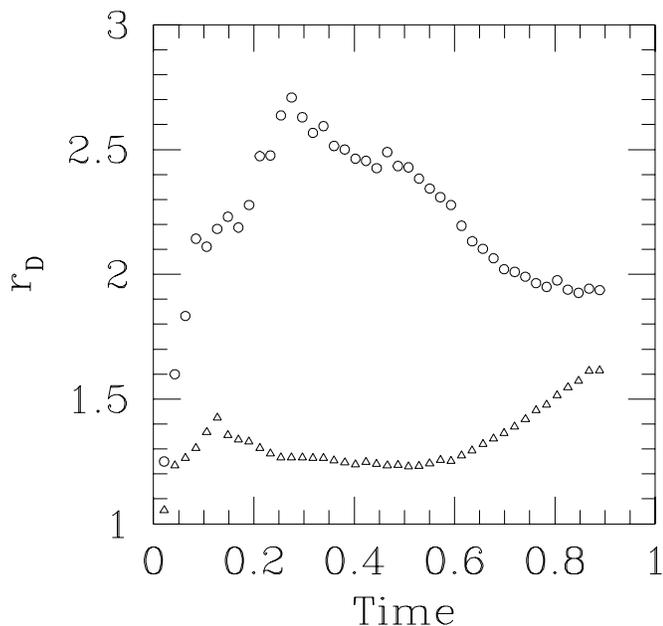,bbllx=53pt,bblly=200pt,bburx=375pt,bbury=504pt,width=3.5in,clip=}
\caption{The arm-length ratios, $r_D(t)$ defined for both
the bow shock and Mach disk locations of Runs A3 and B3 as functions
of time rescaled to units of Myr. The triangles
denote the bow shock ratio and the circles denote the Mach
disk ratio. \label{3Dratio}
}
\end{figure}

Figure~\ref{3Dratio} shows the arm-length ratio, $r_D(t)$ for B3 to A3; this is plotted
using both the bow-shock locations and the Mach disk locations.  We see that 
the Mach-disk ratio basically follows the trend we would expect, with a substantial
increase at early times while the faster jet escapes the denser core and
enters the ISM, and then a more gradual decline after both jets reach
the ISM.  Unfortunately, our simulations could not be followed long enough to
see the ratio continue to decline once both jets were in the constant density
ICM, as the faster one leaves the grid before the slower one gets to that second
interface (Fig.~\ref{3Dratio}).  The bow-shock ratio is less variable, but,
after first rising, and then slightly declining, it again rises at moderate times
as the Mach disk in B3 catches up with the bow shock, while the Mach disk in A3
is only starting to do so.  These simulations can be scaled so 
as to fit FR II
radio sources. For a temperature of the external gas of about
$10^6$K, the velocity of the jet with respect to the ambient medium
is $0.4c$ and the power of the
jet is $L_b \sim 7\times10^{46}$  erg s$^{-1}$ for a central ambient density
of  $n_0 = 1$ cm$^{-3}$.  We have used this value of $L_b$ in our analytical
calculations (\S 2 and 3) for scaling the time units.

\section{Discussion and Conclusions}
The asymmetric models considered above can explain the observed trend 
in the  asymmetry parameters $r_D$ and $r_L$ 
(Saikia et al.\ 1995, 2003; Arshakian \& Longair 2000).
Our model for  jets in an asymmetric environment oriented in the 
plane of the sky explains the large number of galaxies and quasars
having $r_D > 1$ and $r_L < 1$. 
Our simulations of the  jets
propagating in asymmetric environments show a similar trend in arm-length
asymmetry as that produced in our analytical calculations.
We note that  the observed ranges of $r_D$ and $r_L$ can be
easily accommodated with the kind of asymmetric density profile discussed here.
The inclusion of time-delay and Doppler boosting effects can explain
the rarer sources with $r_L > 1$, and the fact that these are relatively
more common for CSS quasars (presumably seen at smaller $\varphi$) than
for CSS radio galaxies.

Although our analytical model for radio source evolution is certainly oversimplified, 
somewhat more sophisticated models (e.g.\ Kaiser, Dennett-Thorpe
\& Alexander 1997; Blundell et al.\ 1999;
 Manolakou \& Kirk 2002; Barai et al.\ 2004), yield similar trends for distances and radio powers
with jet power, external medium density and time.  Larger differences emerge
at late times and at high redshifts, when different treatments of adiabatic
losses and our neglect of inverse Compton losses become important; however,
these are not relevant for the young CSS sources, since the synchrotron
losses from these compact lobes will always exceed the inverse Compton
losses (e.g., Blundell et al.\ 1999).  The key advantage our
simple model has over those more complicated approaches is that we avoid
the necessity of
performing numerical integrals to obtain values for $L_R$, so the estimates
of Sects.\ 2 and 3 are thus possible.

Although the analytical and numerical models agree in a qualitative sense,
it is not surprising that there are differences between the analytical and the numerical
calculations for the arm-length ratio. 
Both the 2-D and 3-D numerical simulations show a 
peak value for $r_D$ of about 2.7. At later times the numerically estimated
$r_D$'s  seem to approach 
asymptotic values greater than 1, although the simulations could not be
continued long enough to confirm this. The analytical maximum $r_D$ is
about 1.7.   
The time taken to reach the first interface is similar in both analytical
and numerical calculations, but in the simulations the second interface
is reached later compared to the times they are reached in the
analytical calculations.  This implies that 
even though the calculated and simulated velocities are similar at early times, 
the velocity of 
the numerical jet, particularly for the 3-D case, is slower than our 
simple ram pressure balance calculations would indicate.
These lower velocities in the numerical simulations can be
understood because the effective working surface in the simulations
is somewhat larger than the analytical model's jet width,
particularly at early times.  These slower speeds,
which are more pronounced for the jet propagating
through the denser gas, also explain most
of the discrepancies between the maximum values attained by $r_D$,
and the longer times needed for $r_D$ to decrease towards unity.

A possible reason why the jet in the 3-D simulation slows even more with respect 
to the analytical estimate is that the simulated jet becomes  elliptical and rotates 
about its own axis, as expected from theoretical calculations
of the fastest growing perturbations (e.g.\ Hardee, Clarke \& Howell 1995).
The jet eventually produces an oblique
 Mach disk, effectively  making the jet thrust smaller while narrowing
the transverse width of the Mach disk 
(e.g.\ Norman 1996; Hooda \& Wiita 1996). Thus, allowing 
for non-axisymmetry in the computations
basically acts to preserve the asymmetry
in $r_D$, and could explain  the larger asymmetry seen in these simulations as
well as the slower velocities at later times, as compared to the 
simplified analytical calculations.

Recently Carvalho \& O'Dea (2002a,b) compared the propagation of jets
of many different values of $\zeta$ and $M_j$.  These simulations also 
used the ZEUS 2-D code and were restricted to axisymmetry, but they
considered both propagation through constant density media and declining
power laws.
Although several parameters are different from our study, we made rough estimates 
of arm-length ratios using their high $M_j$ simulations through an ambient 
medium of constant density 
($\delta=0.0$), and declining density ($\delta=0.75$), 
at a time of 0.5 Myr;  $r_D$ is  
about 2.0, 1.3 and 1.1 for $\zeta$ of 0.0276, 0.0542 and 0.1106, respectively. 
The ranges of the observed $r_D$ thus argue for a small $\zeta (< 0.01)$ for 
extragalactic radio jets. 
Such low $\zeta$ values have long been known to be 
favoured by the observed widths of cocoons seen in FR II radio sources 
(Norman et al.\ 1982). 

Carvalho (1998) performed analytical calculations of jets 
``scattering'' or penetrating
through clouds in an otherwise uniform medium; the asymmetry is produced 
due to the statistical variance in the 
 number of clouds the jets cross through. The arm-length 
asymmetry produced in his model decreases as the jet propagates
forward but the symmetrization occurs at much smaller scales. 
To explain the corresponding flux density ratios would require the shorter
arm to be colliding with a cloud at the time of observation or at least
the presence of more contrast in
densities between the clouds and regular ISM on the side with the  shorter  arm 
(Saikia et al.\ 1995, 2002).   In our simulations  the asymmetry
builds up  as time progresses before getting symmetrized at much later times
with the corresponding flux density ratio changing in the correct sense. 

Recent 3-D simulations  of 
jet/cloud interactions typically either show the jets destroying the clouds
or being halted by them; for a small range of jet/cloud parameters,
some  highly distorted structures and 
deflected arms can be produced (Higgins et al.\ 1999; Wang et al.\
2000).  
Such simulations may be applicable to only a small number of CSSs.
These cloudy medium models can confine the 
CSS sources to sub-galactic sizes with ages of the order of the larger
radio sources (Carvalho 1998). But most observational evidence
(Fanti et al.\ 1995; Readhead et al.\ 1996a,b; Pihlstr{\"o}m et al.\ 2003) 
suggests that the CSS sources are young, not ``frustrated''. Our model
predictions fit the  
observational studies connecting CSS sources with both youth and
higher degrees of asymmetry.

An analysis of the breaks in the spectra of 50 CSS sources, which
included many measurements at 230 GHz, supports the picture that
they are very young, with most having estimated ages of $< 10^5$yr 
(Murgia et al.\ 1999).  Our models correspond to
ages in the observationally estimated ranges of $<10^5$--$10^6$yr.
Our models also have the key advantage of producing the types of asymmetries in 
arm-lengths and luminosities that are characteristic of smaller sources.

\acknowledgements
We thank Gopal-Krishna and Kandaswamy Subramanian for their comments on
the manuscript, and Srianand for his help.  We are grateful to the
anonymous referee for finding an error in the original manuscript and for 
suggestions that have improved its clarity.
The two-dimensional numerical simulations were performed on the Pittsburgh
Supercomputer Center Cray C90, under an award to PJW funded by the NSF. 
SJ thanks Georgia State University for hospitality and PJW is grateful 
for hospitality at NCRA, RRI, and Princeton University Observatory.  This work
was supported in part by NASA grant NAG 5-3098 and by RPE and 
Strategic Initiative Funds to PEGA at GSU.

\end{document}